# Monolayer $MoS_2$/GaAs heterostructure self-driven photodetector with extremely high detectivity

Zhijuan Xu, † Shisheng Lin,*†‡ Xiaoqiang Li, † Shengjiao Zhang, † Zhiqian Wu, † Wenli Xu, † Yanghua Lu, † Sen Xu †

†Department of Information Science and Electronic Engineering, Zhejiang University, Hangzhou, 310027, China

‡ State Key Laboratory of Modern Optical Instrumentation, Zhejiang University, Hangzhou, 310027, China

*Corresponding author. Tel: +86-0571-87951555, Email: shishenglin@zju.edu.cn.

**Abstract:** Two dimensional material/semiconductor heterostructures offer alternative platforms for optoelectronic devices other than conventional Schottky and p-n junction devices. Herein, we use $MoS_2$/GaAs heterojunction as a self-driven photodetector with wide response band width from ultraviolet to visible light, which exhibits high sensitivity to the incident light of 635 nm with responsivity as 446 mA/W and detectivity as $5.9 \times 10^{13}$ Jones (Jones = cm $Hz^{1/2}$ $W^{-1}$), respectively. Employing interface design by inserting h-BN and photo-induced doping by covering Si quantum dots on the device, the responsivity is increased to 419 mA/W for incident light of 635 nm. Distinctly, attributing to the low dark current of the $MoS_2$/h-BN/GaAs sandwich structure based on the self-driven operation condition, the detectivity shows extremely high value of $1.9 \times 10^{14}$ Jones for incident light of 635 nm, which is higher than all the reported values of the $MoS_2$ based photodetectors. Further investigations reveal that the $MoS_2$/GaAs based photodetectors have response speed with the typical rise/fall time as 17/31 μs. The photodetectors are stable while sealed with polymethyl methacrylate after storage in air for one month. These results imply that monolayer $MoS_2$/GaAs heterojunction may have great potential for practical applications as high performance self-driven photodetectors.

**KEYWORDS:** Molybdenum disulfide, GaAs, heterostructure, photodetector.

## Introduction

Since the discovery of graphene in the year of 2004, capabilities of modern electronic and optoelectronic devices are being extended by the emerging field of two-dimensional (2D) materials [1-4]. Photodetector is one of the important optoelectronic devices which are widely used in various fields including military applications and commercial products in everyday lives. Research of 2D materials based photodetector is a hot topic based on the outstanding properties of the 2D materials [5-8]. Graphene based photodetectors have reported to have responsivity as high as ~$10^7$ A/W through the enhanced light absorption with covering semiconductor quantum dots [9]. Using similar technique of quantum dots enhanced photon absorption, $MoS_2$ based photodetector achieves responsivity as high as $6\times10^5$ A/W [10]. Detectivity of a photodetector reflects the sensitivity of the device, which is critical for the detector to detect weak signals. The abovementioned transistor based photodetectors have high responsivities while the detectivities are not as good since the external bias cause relative high dark current. A photodiode type photodetector is a semiconductor junction that converts light into current. The current is generated when photons are absorbed in the photodiode. A photodiode can work with a bias voltage or zero external bias (self-driven) mode. When a photodetector works in a zero external bias mode, we call it as self-driven photodetector in this paper. [11, 12] Among various types of photodetectors, photodiode-based self-driven photodetectors have the special advantage of high detectivity for weak signals based on the low dark current as no external bias voltage is needed for the self-driven operation condition. Besides, photodetectors of this type possess merits of saving energy, small device size and suitable for applications in extreme conditions [13-15]. Until now, kinds of photodiodes including p–n junctions and Schottky junctions have been widely used as photodetectors [16, 17]. For conventional p-n junction photodetectors, the junction locates beneath the front surface, which causes recombination loss as the carriers generated at the front surface need diffuse to the junction to be separated and collected. For the conventional metal/semiconductor Schottky junction photodetectors, the front metal introduces unexpected reflection of the incident light signal. 2D materials/semiconductor heterojunction behaves similarly as Schottky junction or p-n junction [18-20]. Moreover, atomic thin nature of the 2D materials in the photodiode devices guarantees directly generation, separation and collection of the carriers as the junction locates at the front surface of the device. Thus, systematically investigation of the 2D materials based self-driven photodetector is highly need for achieving high performance photodetector devices considering the outstanding electrical and optical properties of the 2D materials [21-23]. Monolayer graphene/germanium self-driven photodetector has been reported to have responsivity and detectivity as 51.8 mA/W and $1.38 \times 10^{10}$ Jones (Jones = cm $Hz^{1/2}$ $W^{−1}$), respectively [24]. However, the zero band gap of graphene usually results in the large dark current of graphene/semiconductor heterostructure, which leads the relatively low detectivity.

Single layer 2D molybdenum disulfide ($MoS_2$) is a semiconductor with a direct band gap of 1.8eV [25]. Compared with other 2D semiconductors, $MoS_2$ can be large-area synthesized with chemical vapor deposition method [26-28]. Thus, $MoS_2$ based photodetectors are promising for low cost practical applications. Moreover, piezoelectric properties of $MoS_2$ allow multi-functional photodetector device design [29]. However, the absence of high-quality p–n junctions makes 2D $MoS_2$-based photodetectors suffering from low detectivity ($10^8$–$10^{10}$ Jones) (Jones = cm $Hz^{1/2}$ $W^{-1}$) and slow response speed (70–$10^6$ μs) [30-32]. The thinnest $MoS_2$/graphene and $WS_2$/$MoS_2$ stacked heterostructure shows a power conversion efficiencies of less than 1% as very small part of light can be absorbed [33, 34]. The monolayer $MoS_2$ itself is too thin to absorb enough incoming light. In the previous paper, we have designed $MoS_2$/GaAs solar cells with a power conversion efficiency over 9.03% [35], which utilizes the good optoelectronic characteristics and high carrier mobility of GaAs [36-38]. In this communication, we design a simple and high performance self-driven photodetector by transferring monolayer $MoS_2$ onto GaAs, which can detect photons from ultraviolet to visible light. The responsivity and detectivity reach 321 mA/W and $3.5 \times 10^{13}$ Jones for the incident light with wavelength of 635 nm, respectively. The typical rise/fall time are as short as 17/31 μs. h-BN is a 2D dielectric material with a band gap of 5.9 eV and dielectric constant of 4.0. By inserting 5 layers of h-BN and forming $MoS_2$/h-BN/GaAs sandwich heterostructure, the dark current of the device is decreased by about one magnitude. Meanwhile, photo-induced doping of MoS$_2$ by Si quantum dots (QDs) are introduced into this device, the responsivity for the photons with wavelength of 635 nm is increased by about 30% from 321 mA/W to 419 mA/W, and the final detectivity is increased into $1.9 \times 10^{14}$ Jones, which is the highest in the reported MoS$_2$ based photodetectors. The stability of the Si QDs/MoS$_2$/h-BN/GaAs photodetector is stable over one month when sealed with polymethyl methacrylate (PMMA).

**Experiments**

Single layer MoS$_2$ film was grown on Si/SiO$_2$ substrate in a quartz tube at 650 $^o$C with CVD method [39]. MoO$_3$ powder and sulfur powder (99.9%, both bought from Aladdin) was used as the precursor. Monolayer h-BN was grown also on copper substrate with CVD technique using B$_3$N$_3$H$_6$ as the precursor at 1000 $^o$C for 30 min [40]. The GaAs substrate is heavily n-type doped with an electron concentration around $1 \times 10^{18}$ cm$^{-3}$. GaAs substrate was cleaned by dipping the samples into 10%wt HCl solution for 5min followed with DI water rinse. Rear Au contact with a thickness of 60nm was thermally evaporated on back surface of GaAs. Then surface passivation of GaAs was realized by remote NH$_3$ plasma treatment for 5min with 120 Watt 27.5 MHz RF generator. After PMMA spun-on, MoS$_2$ on Si/SiO$_2$ substrate was immersed into deionized water to lift-off the PMMA-MoS$_2$ films. After transferring, PMMA was removed by immersing the samples into acetone for 20 min. h-BN was transferred onto GaAs substrate also with the supporting layer of PMMA and multilayer h-BN was achieved by multiple transfer of monolayer h-BN. MoS$_2$ was characterized by Raman spectroscopy (Renishaw inVia Reflex) with the excitation wavelength of 532 nm. The current-voltage characteristics of $MoS_2$/GaAs photodetector devices were tested by Agilent B1500A system. Photo response measurements were carried out using pulse lasers with wavelengths of 325 nm, 405 nm and 635 nm at 1 MHz repetition rate and 50 ps pulse duration. The diameter of the excitation laser spot was 1 mm. The photocurrent was recorded with Keithley 2400 Source Meter. The electrical response was recorded with an oscilloscope (Tektronix, TDS2012B). The incident power into the device was calculated from the measured laser power, the laser spot size and the active area of the device. Spectra response of the photodetector devices were measured with PV Measurements QEXL system. To avoid unexpected oxidation of the monolayer $MoS_2$, the devices were sealed by PMMA through simply spinning-on PMMA on the top of the photodetector device followed with 120 $^o$C / 2 min curing process on a hot plate. Then the devices were stored in normal indoor environment and the stability was checked using the same characterization method used before.

## Results and discussion

Fig. 1a shows the schematic structure of the $MoS_2$/GaAs heterojunction photodetector, which is composed of rear Au contact, GaAs substrate, $MoS_2$ layer and the front Au contact from bottom to top. The outer size of the Au contact area is roughly 1.5 mm × 1.5 mm. The active area (0.5mm×0.5mm) of the device is defined with the opened window in the front Au contact locating at the center of the Au contact area. Raman spectrum of the $MoS_2$ on Si/$SiO_2$ substrate is shown in Fig. 1b.

The Raman peaks corresponding to $E^1_{2g}$ and $A_{1g}$ modes of $MoS_2$ locate at 383.4 cm$^{-1}$ and 404.6 cm$^{-1}$, respectively, indicating the monolayer nature of $MoS_2$ [41]. Fig. 1c shows the schematic fabrication processes and optical microscopy (OM) image of $MoS_2$/GaAs heterojunction based photodetector. The processes as sequence are removal of the native oxide on the GaAs substrate, Au rear contact evaporation, clean and passivation of GaAs front surface, $MoS_2$ transferring, lithography process for front contact mask, front contact deposition and the final lift-off process. As mentioned above, the spot size of laser is 1 mm in diameter. During measurements of the photoresponse, lasers are aimed to the center of the active area. The laser spot is larger than the active area, which guarantees the whole active area is illuminated.

The schematic electronic band structure of the $MoS_2$/GaAs heterojunction is shown in Fig. 2a. The electron affinity of GaAs ($\chi_{GA}$) is 4.07eV. Fermi level of GaAs ($E_{F-GA}$) used in this study locates near the bottom of the conduction band ($E_{C-GA}$) as the n-type doping concentration is around $10^{18}$ cm$^{-3}$. The band gap of monolayer $MoS_2$ is 1.8 eV, and the electron affinity (energy gap between vacuum level and the bottom level of conduction band $E_{C-MS}$) of $MoS_2$ ($\chi_{MS}$) is 4.0 eV [42]. As the grown $MoS_2$ is not intentionally doped, Fermi level of $MoS_2$ ($E_{F-MS}$) locates near the middle position of the band gap. When forming the $MoS_2$/GaAs heterojunction, some amount of electrons transfer from n-type GaAs into $MoS_2$ because of the Fermi level difference, which leads to upward bending of the electronic bands of GaAs at the $MoS_2$/GaAs interface and hence a built-in harrier is formed. When this heterojunction is under illumination, excess holes and electrons are collected in $MoS_2$ and GaAs respectively.

The dark current density-voltage (*J-V*) curves of the MoS$_2$/GaAs shown in Fig. 2b shows good rectifying characteristics. Inset of Fig. 2b shows the reverse current at voltages from 0 V to -0.5 V, where the dark current at zero bias is as low as $4.5 \times 10^{-13}$ A. Meanwhile, *J-V* curve of the MoS$_2$/GaAs heterostructure photodetector under AM1.5G illumination also shown in Fig. 2b shows an open-circuit voltage of 0.55 V, implying the good photovoltaic properties of the heterojunction. Based on the photovoltaic properties and the atomic thin top MoS$_2$ layer, this MoS$_2$/GaAs heterostructure device can work as self-driven photodetector with good response as incident photon can be directly and efficiently converted into electricity. Fig. 2c displays the photocurrent of the device under laser illumination with different wavelength and intensities. It is noteworthy that these results are obtained in the self-driven condition and no bias voltage is applied on the device. Photocurrent increases linearly with the laser power for the lasers with different wavelength values, and the larger the wavelength, the higher the photocurrent, which is caused by more incident photons of the laser with larger wavelength. Fig. 2d shows the time response when the MoS$_2$/GaAs photodetector exposed to the lasers with different wavelength, where identical time response can be found. The response speed of the photodetector is characterized with the response time ($\tau_r$), which is the time interval for the response to rise from 10% to 90% of its peak value. The recovery time ($\tau_f$) is the time interval for the response to decay from 90% to 10% of its peak. The $\tau_r$ and $\tau_f$ are 17 μs and 31 μs, respectively. Series resistances (R) of the device can be fitted from the current-voltage curve and the typical value is 630 Ω. The capacitance (C) of the device measured using Agilent B1500A system is 3.27 pF. The RC time constant is $2.1 \times 10^{-9}$ sec and the cut-off frequency is calculated as $7.6 \times 10^{7}$ Hz. The typical rise time and fall time is 17 μs and 31 μs, which are about four magnitudes slower than the RC time constant, which suggests that the response speed of the $MoS_2$/GaAs can be further improved in the future work including designing smaller device size and optimizing the contact pattern.

Responsivity and detectivity are the main parameters for a photodetector device. Responsivity represents the photocurrent generated per unit of the incident power by the device, while detectivity stands for the ability for sensing weak signal of the photodetector. Responsivity can be determined with equation (1) shown below:

$$Responsivity = \frac{I_L - I_D}{P_{laser}} \qquad (1)$$

where $I_L$ is the current under laser illumination, $I_D$ is the dark current, $P_{laser}$ is the power of the incident laser. The noise in a photodiode can take two forms. The first is the shot noise of the dark current, which results from the statistical uncertainty in photon arrival rate from the view of optics or the discrete nature of electric charge collected by the electrodes from the view of electronics. The value of the shot noise is [43]:

$$I_{shot} = \sqrt{2qI_D} \qquad (2)$$

where q is the charge of electron, $I_D$ is the dark current of the photodetector. The value of $I_{shot}$ is dependent on the value of $I_D$. Reduced $I_D$ helps to obtain low shot noise. The second noise source is the thermal noise, also known as Johnson noise, which is caused by the thermal fluctuations of stationary charge carriers and can be expressed

as [44]:

$$I_{Johnson} = \sqrt{\frac{4kT}{R_{shunt}}} \quad (3)$$

where k is the Boltzmann constant, T is the absolute temperature and $R_{shunt}$ is the shunt resistance of the photodiode. $R_{shunt}$ reflects the leakage properties under reverse bias of the photodiode and can be obtained through the dark current-voltage curve by linearly fitting the reverse current at low reverse voltage range. Based on the obtained $R_{shunt}$ and $I_D$, the values of shot noise and Johnson noise are $9.1 \times 10^{-16}$ A and $1.1 \times 10^{-15}$ A, respectively. Here we found Johnson noise is slightly higher than shot noise.

Thus, detectivity can be determined with equations (4) shown below.

$$Detectivity = \frac{\sqrt{A} \cdot R}{\sqrt{\frac{4kT}{R_{shunt}}}} \quad (4)$$

where $A$ is the active area of the device. Fig. 3a shows the obtained values of responsivity of the $MoS_2$/GaAs photodetector. Besides self-driven condition, the responsivity values with different reverse bias voltages are also plotted in Fig. 3a. Under self-driven condition, the responsivity values for the lasers with wavelengths of 325nm, 405 nm and 635 nm are 96.5 mA/W, 160 mA/W and 321 mA/W, respectively. When applying reverse bias, the photocurrent is increased, leading to the roughly linear enhancement of the responsivity. For example, reverse bias voltage of -1.0 V increases the responsivity up to 141 mA/W, 247 mA/W and 472 mA/W for the device illuminated by lasers with wavelengths of 325 nm, 405 nm and 635 nm, respectively. Fig. 3b shows the detectivity of the photodetector. Corresponding to the responsivity values, detectivity values under illumination of 635 nm laser are the highest based on different reverse bias voltages. The highest detectivity value is $3.5 \times 10^{13}$ Jones with zero bias voltage. The highest detectivity values for the conditions illuminated with 325 nm and 405 nm lasers are $1.2 \times 10^{13}$ Jones and $1.7 \times 10^{13}$ Jones, respectively.

As mentioned above, the key parameters for photodetectors are responsivity and detectivity. Based on the simple planar device structure and physical picture of the $MoS_2$/GaAs heterojunction, much room exists for improving the performance of the photodetector. Herein, photo-induced doping is adopted for improving the responsivity and interface design are employed for enhancing detectivity in the device as shown in Fig. 4a. Different from the structure shown in Fig. 1a, Si QDs is placed on the surface of $MoS_2$ acting as a photo-induced dopant and 5 layers of h-BN are inserting into the interface of $MoS_2$/GaAs heterojunction. Photo-induced doping can increase the current of the graphene/GaAs heterostructure solar cell as we demonstrated in our former work [45]. Thus enhanced responsivity can be expected similarly in the $MoS_2$/GaAs photodetector. As the 2D semiconductor $MoS_2$ is atomic thin, the charge transfer from GaAs to $MoS_2$ shifts the Fermi level of $MoS_2$ and thus reduces the barrier of the junction. The larger the barrier height, the lower the dark current of the heterojunction. As we have reported, inserting h-BN can suppress the charge transfer in the $MoS_2$/GaAs heterojunction, thus higher barrier height and $R_{shunt}$ can be obtained and hence thermal noise can be reduced. Hence, higher detectivity can be expected. Fig. 4b shows the electronic band structure and charge transfer process of the Si QDs doped $MoS_2$/h-BN/GaAs photodetector. The valence and conduction bands of both Si QDs ($E_{C-Si}$ and $E_{V-Si}$) and GaAs ($E_{C-GaAs}$ and $E_{V-GaAs}$) bend upward near MoS$_2$ layer to maintain Fermi level balance. Based on the electronic band alignment, electron transfer from GaAs to MoS$_2$ can be suppressed during the formation of the heterojunction while transport of excess holes under illumination from GaAs to MoS$_2$ is almost unaffected after inserting the 2D h-BN layer，which means that barrier height can be increased while the photocurrent under illumination is almost not influenced. Under illumination, excess electrons and holes generated in GaAs are collected by GaAs and MoS$_2$ respectively. Moreover, excess holes generated in Si QDs can hop into MoS$_2$ directly and excess electrons stay in the Si QDs, which results in negative charged Si QDs. The injected holes and the field from this negative charged QDs lead to p-type doping of thin MoS$_2$ layer, which results in the even higher barrier height and higher collection efficient of photo-generated current. Fig. 4c shows the *J-V* curves of the MoS$_2$/GaAs, MoS$_2$/h-BN/GaAs and Si QDs/MoS$_2$/h-BN/GaAs heterostructures under AM1.5G illumination. Compared with MoS$_2$/GaAs heterojunction, open-circuit voltage of MoS$_2$/h-BN/GaAs junction is increased from 0.58 V to 0.62 V by increasing the barrier height, while the short-circuit current density stays almost unchanged. After covering Si QDs on the MoS$_2$/h-BN/GaAs junction and forming Si QDs/MoS$_2$/h-BN/GaAs heterostructure, short circuit current is increased from 17.2 mA/cm$^2$ to 22.7 mA/cm$^2$. Meanwhile, open circuit voltage is slightly increased from 0.62 V to 0.63 V. The increase of the short-circuit current density can be analyzed from the comparison of spectral response measured from MoS$_2$/h-BN/GaAs and Si

QDs/MoS$_2$/h-BN/GaAs heterostructures under AM1.5G illumination, which can be seen in Fig. 4d. The enhancement of the spectra response mainly locates in the wavelength range of 400 nm to 800 nm, as the photons with short wavelength can absorbed near the surface in the MoS$_2$/GaAs heterostructure where the collection efficiency is high thus photo-induced doping from Si QDs is not as effective as for the photons with middle wavelength and the photons with wavelength near the band gap edge cannot generate excess carriers effectively in the Si QDs. The band selectivity of the MoS$_2$/GaAs heterostructure based photodetector is mainly determined by the light absorption in the GaAs substrate as very small part of incident light is absorbed in ultrathin top MoS$_2$ layer. The cut-off wavelength is around 873 nm, as shown in Fig. 4d, determined by the band gap of GaAs. The capability for detecting longer wavelength photons can be achieved by using substrate with smaller band gap.

Fig. 5a shows the responsivity of the Si QDs/MoS$_2$/h-BN/GaAs photodetector with different bias voltages. Under zero bias, the responsivities for the lasers with wavelengths of 325 nm, 405 nm and 635 nm are 105 mA/W, 182 mA/W and 419 mA/W, respectively. Similar with the MoS$_2$/GaAs device, when applying reverse bias of -1.0 V, the responsivities are increased with the values up to 198 mA/W, 319 mA/W and 657 mA/W for the conditions illuminated by lasers with wavelengths of 325 nm, 405 nm and 635 nm, respectively. It is noteworthy that even larger responsivity can be obtained with larger reversing bias voltage, while the detectivity will be decreased as the dark current increases. Fig. 5b shows the detectivity of the Si QDs/MoS$_2$/h-BN/GaAs photodetector. The R$_{shunt}$ of the device with interface h-BN is increased by from $1.4 \times 10^{10}$ Ω to $3.9 \times 10^{10}$ Ω. Correspondingly, the detectivity is increased based on equation (4). Similarly, detectivity values under illumination of 635 nm laser are the highest based on different reverse bias voltages. The highest detectivity value is increased up to $1.9 \times 10^{14}$ Jones with zero bias voltage. The highest detectivity values for the conditions illuminated by lasers with wavelengths of 325 nm and 405 nm are $3.9 \times 10^{13}$ Jones and $6.5 \times 10^{13}$ Jones, respectively. The obtained detectivity values are the highest values among all the reported results of $MoS_2$ based photodetectors [43, 46-48]. The possible drawback of the $MoS_2$ based device is that $MoS_2$ tends to be oxidized in air especially when the lattice of $MoS_2$ is imperfect [49]. However, since monolayer $MoS_2$ is atomic thin, we can finely protect it through covering thick anti-oxidation layer. Herein, we use PMMA to seal the device to illuminate unexpected oxidation. As shown in Fig. 5c, after storage in air for one month, the photo-response stays unchanged, demonstrating the device is stable for practical application. This result demonstrates that $MoS_2$ based device can be electrical stable if well encapsulated.

## Conclusions

In summary, $MoS_2$/GaAs heterojunction is designed as a high sensitive self-driven photodetector. By inserting h-BN into the interface of $MoS_2$/GaAs heterostructure to decrease the dark current and employing photo-induced doping to increase the photocurrent, the responsivity and the detectivity reach 419 mA/W and $1.9 \times 10^{14}$ Jones, respectively. The obtained detectivity is higher than all the reported values of the $MoS_2$ based photodetectors to now. The response speed of this MoS$_2$/GaAs based photodetector is very quick with typical rise/fall time are 17/31 μs. Meanwhile, the photodetectors are stable with PMMA sealing after storage in air for one month. These results suggest that monolayer MoS$_2$/GaAs heterojunction may have great potential for practical applications as high performance self-driven photodetectors.

**Acknowledgements**

S. S. Lin thanks the support from the National Natural Science Foundation of China (51202216) and X. Q. Li thanks the support from the National Natural Science Foundation of China (51502264) and Postdoctoral Science Foundation of China (2013M540491).

**Figure captions**

Figure 1. (a) Schematic structure of the $MoS_2$/GaAs photodetector. (b) Raman spectrum of the monolayer $MoS_2$. (c) Fabrication process flow and OM image of the $MoS_2$/GaAs photodetector.

Figure 2. (a) Electronic band structure of the $MoS_2$/GaAs heterojunction. (b) Current density-voltage curves of the $MoS_2$/GaAs heterojunction in the dark and under AM1.5G illumination. Inset shows the reverse current at voltages from 0 V to -0.5 V. (c) Photocurrent under laser illumination with different wavelength and power. (d) One typical cycle of photoresponse of the $MoS_2$/GaAs photodetector.

Figure 3. Responsivity values (a) and detectivity values (b) of the $MoS_2$/GaAs photodetector illuminated by lasers with wavelengths of 325 nm, 405 nm and 635 nm.

Figure 4. Schematic structure (a) and electronic band alignment (b) of the Si QDs/$MoS_2$/h-BN/GaAs photodetector. (c) J-V curves of the $MoS_2$/GaAs, $MoS_2$/h-BN/GaAs and Si QDs/$MoS_2$/h-BN/GaAs heterostructures under AM1.5G illumination. (d) Spectra response of $MoS_2$/h-BN/GaAs and Si QDs/$MoS_2$/h-BN/GaAs heterostructures.

Figure 5. Responsivity values (a) and detectivity values (b) of the Si QDs/$MoS_2$/h-BN/GaAs photodetector illuminated under lasers with wavelengths of 325 nm, 405 nm and 635 nm. (c) Normalized response of the as fabricated device and the same device after storage in air for one month.

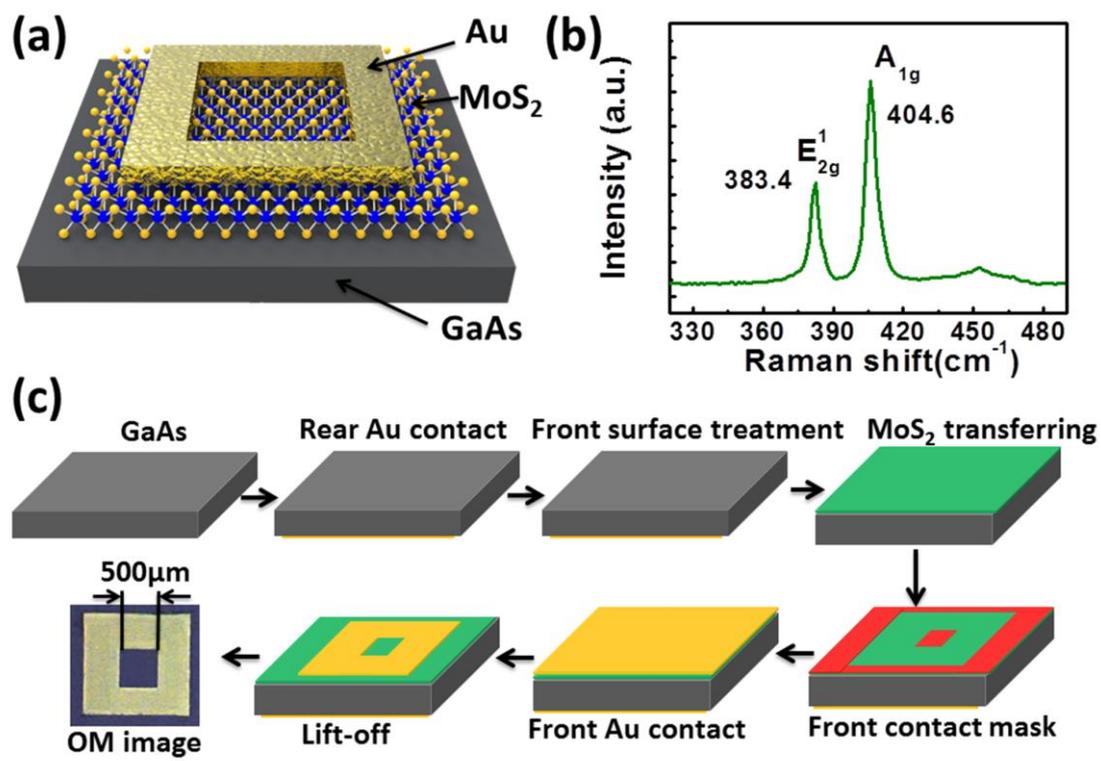

Xu et al., Fig. 1

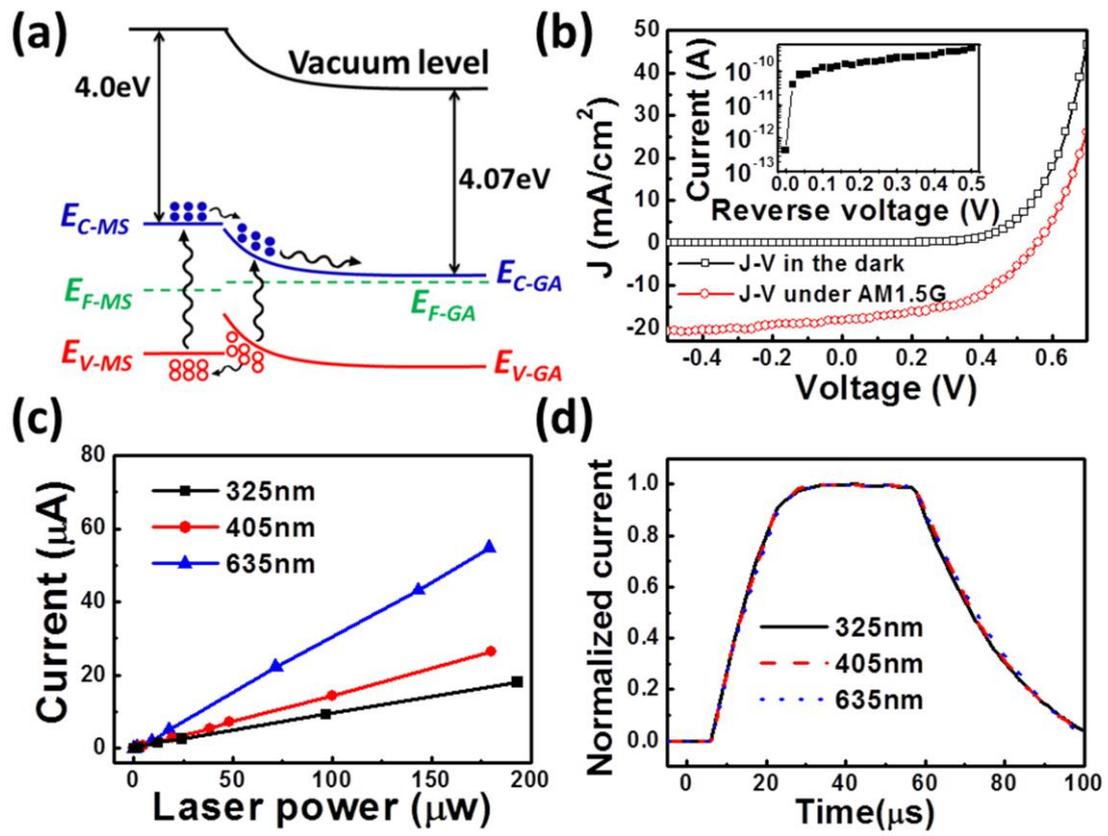

Xu et al., Fig. 2

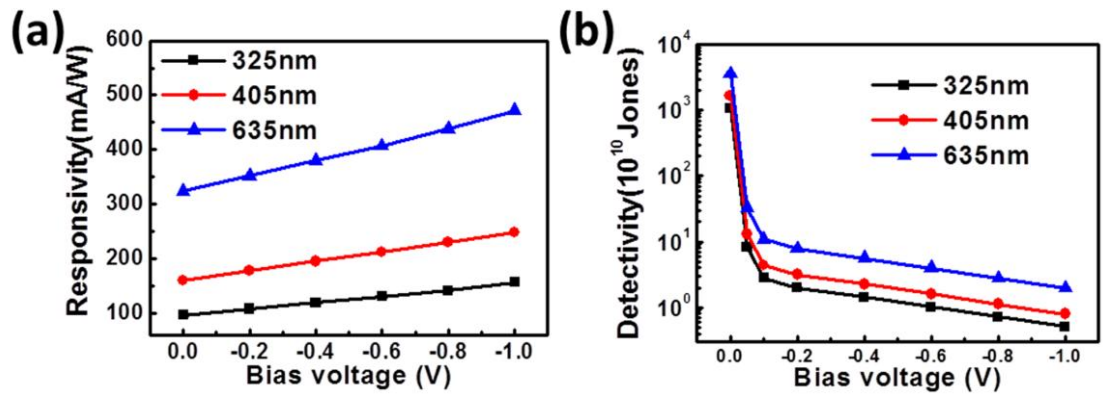

Xu et al., Fig. 3

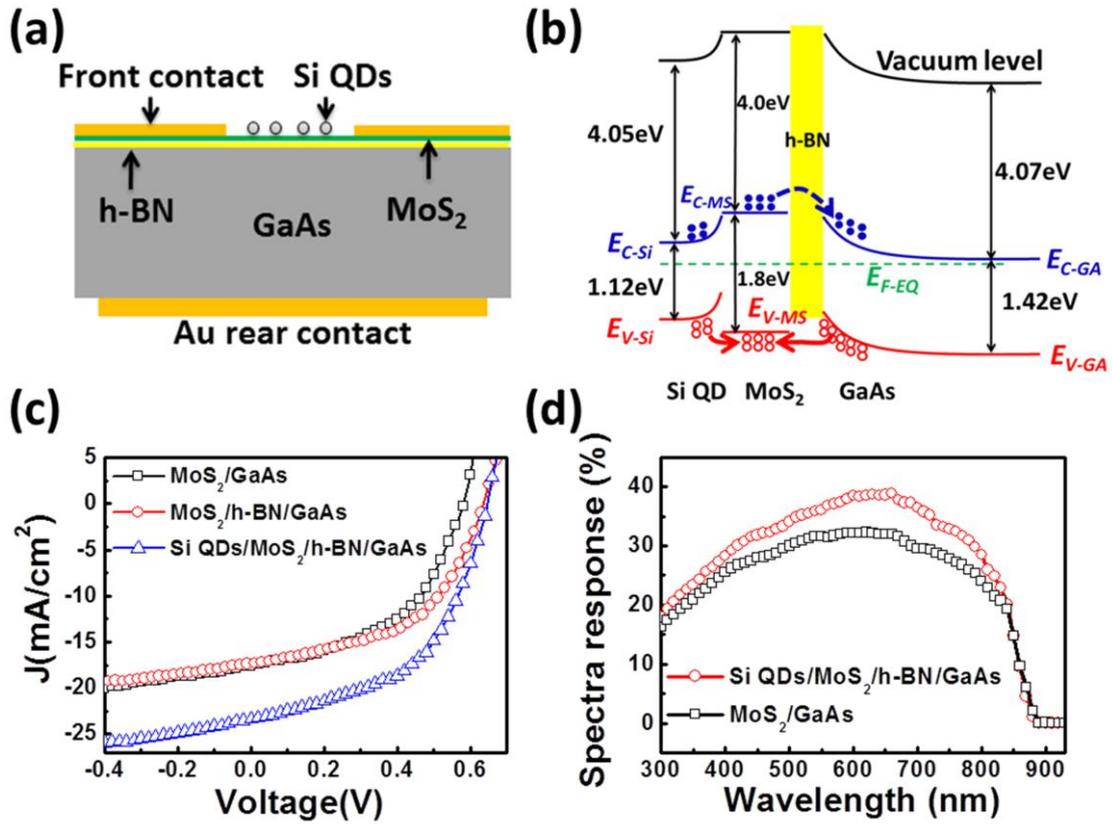

Xu et al., Fig. 4

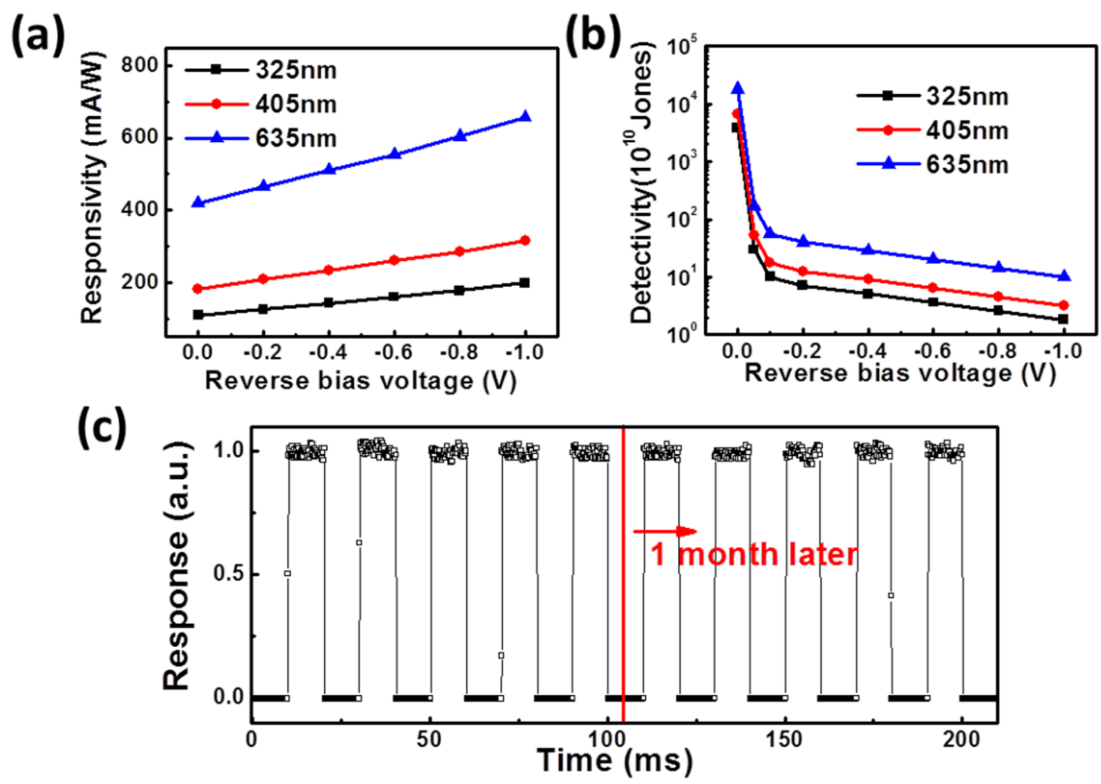

Xu et al., Fig. 5